# Resonant Precession of Magnetization and Precession - Induced DC voltages in FeGaB Thin Films.

**Prabesh Bajracharya, Vinay Sharma, Anthony Johnson, and Ramesh C. Budhani***

**Department of Physics, Morgan State University, Baltimore, MD, 21251 USA**

*Ramesh.budhani@morgan.edu

**Abstract**

Measurements of frequency dependent ferromagnetic resonance (FMR) and spin pumping driven dc voltage ($V_{dc}$) are reported for amorphous films of $Fe_{78}Ga_{13}B_9$ (FeGaB) alloy to address the phenomenon of self-induced inverse spin Hall effect (ISHE) in plain films of metallic ferromagnets. The $V_{dc}$ signal, which is antisymmetric on field reversal, comprises of symmetric and asymmetric Lorentzians centered around the resonance field. Dominant role of thin film size effects is seen in setting the magnitude of static magnetization, $V_{dc}$ and dynamics of magnetization precession in thinner films (≤ 8 nm). The film thickness dependence of magnetization parameters indicates the presence of a magnetically disordered region at the film – substrate interface, which may promote preferential flow of spins generated by the precessing magnetization towards the substrate. However, the $V_{dc}$ signal also draws contributions from rectification effects of a ≈ 0.4 % anisotropic magnetoresistance and a large (≈ 54 nΩ.m) anomalous Hall resistivity (AHR) of these films which ride over the effect of spin – orbit coupling driven spin-to-charge conversion near the film – substrate interface. We have addressed these data in the framework of the existing theories of electrodynamics of a ferromagnetic film subjected to radio-frequency field in a coplanar waveguide geometry. Our estimation of the self-induced ISHE for the sample with 54 nΩ.m AHR shows that it may contribute significantly (≈ 90%) to the measured symmetric voltage. This study is expected to be very useful for fully understanding the spin pumping induced dc voltages in metallic ferromagnets with disordered interfaces and large anomalous Hall effect.

# I. Introduction

The observation of an inverse spin Hall effect (ISHE) – like dc voltage in plain films of metallic ferromagnets (FM) at the ferromagnetic resonance (FMR) frequency has generated considerable interest in recent years [1-6]. The ISHE is a process by which spin currents injected from a ferromagnet into a proximate non-magnetic metal (NM) of strong spin-orbit-coupling (SOC) are converted into charge currents [7, 8]. This process is driven by spin Hall effect (SHE) and interfacial Rashba Edelstein effect (REE) [9-11]. While the origin of a dc voltage induced by thermally pumped spin currents is well-understood [12-14], such is not the case for spin pumping by radio frequency (RF) assisted resonant precession of magnetization. Some earlier measurements of ISHE on FM/NM bilayers at resonance have addressed the role of effects such as spin rectification due to a large anisotropic magnetoresistance (AMR) and anomalous Hall effect (AHE), and thermal gradient induced dc voltages in contaminating the true ISHE signal [15-18]. These studies have also reported measurements of a dc voltage at resonance in control samples of plain FM films, and this weak antisymmetric Lorentzian signal has been attributed to AMR induced spin rectification effect (SRE) in the metallic ferromagnet. However, extensive studies of SRE at resonance in thin film of metallic ferromagnets performed in coplanar waveguide as well as cavity spectrometers have shown that the rectified voltage has both field symmetric and field asymmetric contributions, depending on the geometry of the experiment [19, 20]. Notwithstanding these results, several recent measurements of the ISHE – like signal at FMR in thin films of metallic ferromagnets present a rather different picture. The data of Tsukahara et. al. [1] on 10 nm thick film of permalloy (Py) driven to resonance in a cavity spectrometer show both symmetric and antisymmetric Lorentzian line shapes. They attribute the symmetric part of the signal to ISHE generated by the non-zero SOC in Py films. Measurements on plain films of Fe and Co also reveal an ISHE-like symmetric signal at resonance [4]. The work of Azevedo et. al. [2] on Py films of different thicknesses shows a robust and predominantly symmetric signal at resonance in the direction of applied magnetic field. These authors have attributed this signal to the process of magnonic charge pumping (MCP) wherein the spin waves emitted by precessing magnetization generate a charge current in the FM layer of a non-zero SOC [21]. Under open circuit conditions, this current produces an ac voltage across the sample together with a small dc component due to AMR. The MCP results from the Onsager reciprocity theorem as the counterpart of spin torque FMR (ST-FMR), where the spin orbit torque produced by ac currents drive the magnetization into precession. Azevedo et. al. [2] argue that surface oxidation of the Py film produces a Rashba field that breaks the inversion symmetry of the system to result in a symmetric dc signal. Tsukahara et. al. [1] have attributed the symmetric voltage signal in Py films at resonance to spin pumping induced ISHE in the disordered regions of the film near its interface with the substrate. However, the theory of SRE clearly shows that under specific configurations of dc and rf fields with respect to the dc voltage leads both AMR and AHE will contribute to a symmetric rectified signal [20]. Therefore, the limited studies of the so-called self-induced ISHE in plain films of metallic ferromagnets need to be augmented with the twin objectives of: (i) correctly interpreting the ISHE data on metallic FMs bilayered with SOC-NMs and various topological materials. (ii) Understanding the true origin of dc currents in plain films of metallic ferromagnets driven to resonance by ac fields.

Here, we address *self-induced ISHE* at FMR in non-crystalline films of a metallic ferromagnet. This material is derived from the well-known high magnetostriction alloy $Fe_{80}Ga_{20}$ [22, 23]. The addition of boron in $Fe_{80}Ga_{20}$ followed by rapid solidification from liquid or vapor phase results in a non-crystalline material [24, 25]. Thin films of the alloy $Fe_{78}Ga_{13}B_9$ display large magneto-elastic coupling together with magnetic softness characterized by a low Gilbert damping coefficient [24, 26, 27]. The amorphous nature of these films allows us to rule out any possible contributions of crystallographic texture to the rectified dc voltage. We have carried out detailed studies of magnetization dynamics with frequency-dependent ferromagnetic resonance, vibrating sample magnetometry (VSM) and spin pumping induced ISHE measurements on $Fe_{78}Ga_{13}B_9$ films as a function of film thickness. Our FMR and VSM measurements establish the formation of a ≈ 1.2 nm thick magnetic dead layer at the film - substrate interface. While the symmetry and sign of the dc signal suggest FMR-driven spin pumping into this layer, we examine this result in the light of possible symmetric voltages emanating from AMR and AHE. While the ambient temperature AMR in FeGaB films is smaller by a factor of ≈ 5 compared to the AMR in permalloy, a large anomalous Hall resistivity (≈ 54 nΩ m) is seen in these films. On the other hand, the AHE in $Ni_{80}Fe_{20}$ is very small (≈ 0.96 nΩ m) due to competing contributions from the Ni and Fe derived electronic states [28]. Our measurements of the dc signal at resonance in FeGaB films indicate that notwithstanding the dead layer which may promote unidirectional flow of spin current, the AHE related rectification makes a sizable contribution to the dc signal.

## II. Experimental details

FeGaB films of thickness ranging from 2.0 nm to 12.0 nm were grown by DC sputtering of a $Fe_{78}Ga_{13}B_9$ alloy target on epitaxial quality (0001) plane sapphire substrates at room temperature in 5 x $10^{-3}$ torr argon pressure at the rate of 0.03 nm/s. Static magnetization measurements were performed at ambient temperature on ≈ 5 x 5 mm$^2$ samples in a vector VSM (MicroSense Model 10 Mark II). For the measurements of FMR and ISHE, we have used a meander line grounded coplanar waveguide (GCPW)-based FMR spectrometer that operates in the frequency and field ranges of 5 to 31 GHz and 0 to ± 1.5 tesla respectively (Fig. 1(a and b)). The ISHE was measured by connecting a twisted pair of thin copper wires to the ends of 2 x 5 mm$^2$ film with silver paint.

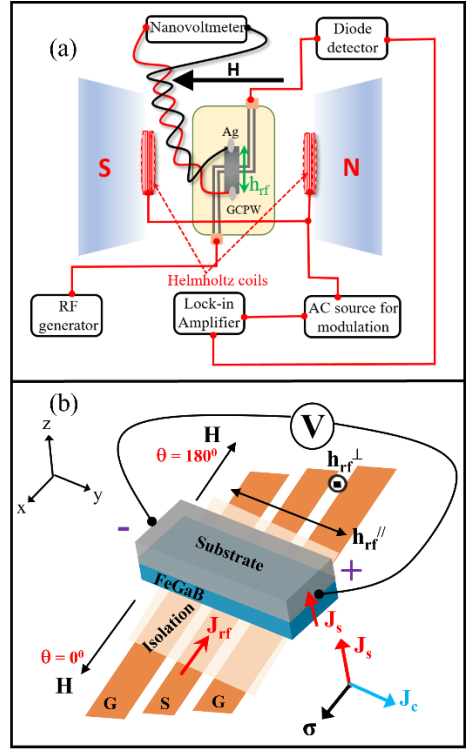

**Figure 1.** **(a)** Schematic illustration of experimental setup used for FMR and ISHE measurements. FMR measurements are performed over a frequency range of 5 to 30 GHz by low amplitude modulation of the dc field at 650 Hz. FeGaB thin film is placed in a flip-chip configuration on the signal line of the grounded coplanar waveguide. A twisted pair of copper wires is used to measure the DC voltage generated across the sample. The DC voltage is measured along y-axis with Y+ contact of the sample connected to the positive terminal of the nanovoltmeter. **(b)** Magnified view of the measurement setup to show the directions of **J**$_s$, **J**$_c$ and spin polarization vector (**σ**), where $J_c =$

$\theta_{SH}(J_s \times \sigma)$. Direction of $\sigma$ and $J_c$ are controlled by changing the direction of dc magnetic field from $\theta = 0^0$ to $180^0$. $J_{rf}$ is the rf current flowing in the cpw signal line

The wired sample was then flip chipped on the section of the waveguide where the rf magnetic field is perpendicular to the external dc magnetic field. A 100 μm thick mica sheet separated the sample and signal line of GCPW for electrical isolation. The dc voltage generated across the sample leads on sweeping the external magnetic field across FMR was measured with a Keithley nanovoltmeter. The FMR signal was recorded on the same sample in a different experiment with phase sensitive detection of absorbed power on low frequency (650 Hz) modulation of the dc field with a set of Helmholtz coils. The radio frequency (RF) source power was changed from -10 dBm to +15 dBm to ensure linear response of the system. The measurements of AMR and planar Hall effect (PHE) were performed by mounting the films, patterned in a Hall bar geometry, on the vertical rotator stage of a physical property measurement system (PPMS). The AHE measurements were performed by mounting the films, patterned in a Hall bar geometry, on the dc resistivity puck of PPMS where dc magnetic field was applied perpendicular to the sample surface.

### III. Results and Discussion
#### A. Static magnetization characteristics

Fig. 2(a) shows the in-plane magnetic hysteresis loops of FeGaB films of different thicknesses ($t_f$) measured at room temperature. These samples have been named as FGB$t_f$, where $t_f$ is the film thickness in nm. The M(H) loops of all samples are characterized by a squareness of low coercive ($\approx 0.1$ mT) indicating a soft magnetic state of in-plane orientation. The saturation magnetization ($\mu_0 M_s$) of these films varies from 0.68 T to 1.43 T on increasing the thickness from 2 nm to 12 nm.

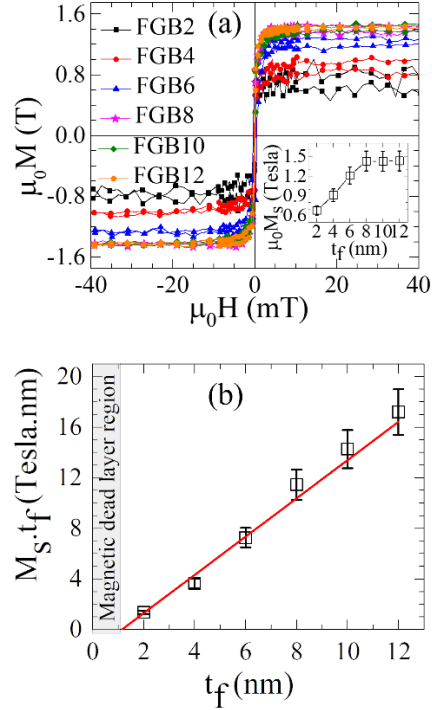

**Figure 2.** (a) dc magnetization loops (M(H)) of FeGaB thin films of different thickness ($t_f$ = 2, 4, 6, 8, 10 and 12 nm) measured at room temperature. Magnetic field is applied parallel to the plane of film. Inset shows saturation magnetization ($\mu_0 M_S$) as a function of film thickness. Thin film size effect is seen prominently in the data for $t_f \leq$ 8nm. (b) Change in the product of $M_S$ and $t_f$ as a function of film thickness. Red line is a linear fit to show the point of magnetic dead layer formation for $t_f \leq 1.2$ nm. Error bars reflect the uncertainty in the measurements of sample volume.

(See supplementary Fig. S1 for anisotropy calculations) A comparison of the M(H) loops shown in Fig. 2(a) with similar data for $Fe_{80}Ga_{20}$ alloy [29, 30] reveals that addition of boron suppresses the magneto-crystalline anisotropy and coercivity of this binary alloy. The saturation magnetization displayed in the inset of Fig. 2(a) increases linearly with the film thickness till $t_f \approx 8$ nm and reaches a constant value beyond this critical thickness. The saturation magnetization of the thicker films ($t_f \geq 8$ nm) agrees with the reported $M_s$ of the films of same composition [25, 27]. In Fig. 2(b) we show the plot of magnetization per unit area ($M_s t_f$) as a function of film thickness. A linear fit to these data suggests the existence of a magnetically dead layer of thickness ≈1.2 nm at the interface.

## B. Dynamic magnetization study using FMR

The results of frequency dependent FMR measurements with external dc field in the plane of the film are shown in Fig. 3(a) and 3(b). The field derivative of the absorbed RF power (dP/dH) at 10 GHz (the FMR line shape) of the 4, 6, 8, 10 and 12 nm thick films is displayed in the inset of Fig. 3(a). The signal from the 2 nm thick film was too weak to detect.

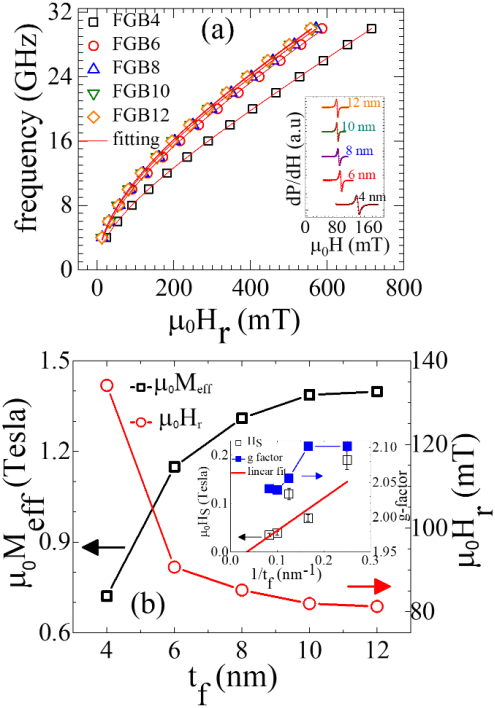

**Figure 3. (a)** Change in FMR resonance field ($\mu_0 H_r$) with microwave frequency for films of varying thickness. Open symbols in the figure are the experimental data calculated from Lorentzian line shape fitting and the solid red line the fit to Kittel equation. Inset shows the differential FMR signal at 10 GHz with Lorentzian line-shape fitting. **(b)** shows the variation of effective saturation magnetization ($\mu_0 M_s$) and FMR resonance field ($\mu_0 H_r$) at 10 GHz as a function of sample thickness. As thickness increases, $\mu_0 M_s$ increases while $\mu_0 H_r$ decreases. Inset shows the induced surface anisotropy field ($H_S$) and g-factor as a function of layer thickness. Red line in the figure is a linear fit to equation $H_s = 4K_s/(M_s t_f)$.

The FMR line shape at all frequencies from 5 to 30 GHz has been analyzed in the framework of a two component Lorentzian function [31]

$$\frac{dP}{dH} = K_1 \frac{4\Delta H(H-H_r)}{[4(H-H_r)^2+(\Delta H)^2]^2} - K_2 \frac{(\Delta H)^2 - 4(H-H_r)^2}{[4(H-H_r)^2+(\Delta H)^2]^2} \quad (1)$$

where $K_1$ and $K_2$ are the symmetric and anti-symmetric coefficients, H is the applied dc field, $H_r$ the resonance field and $\Delta H$ is the full width at half maximum of the resonance. The main panel of Fig. 3(a) displays the variation of $H_r$ as a function of frequency for all five samples.

These data follow the simple Kittel equation [32, 33]

$$f = \frac{\gamma}{2\pi}\mu_0\sqrt{H(H+M_{eff})} \quad (2)$$

for a ferromagnetic film with in-plane magnetization and no preferred anisotropy axis. Here $\mu_0 M_{eff}$ is the effective saturation magnetization that considers surface anisotropies, f is the excitation frequency, $\gamma = \frac{g\mu_B}{\hbar}$ is the gyromagnetic ratio, g is the Lande's g-factor, $\mu_B$ is the Bohr magneton, and ℏ is the Planck's constant. We used $\mu_0 M_{eff}$ and g values as free parameters while fitting the experimental data to Eq. 2. The changes in $\mu_0 M_{eff}$ and $\mu_0 H_r$ (at $f$ = 10 GHz) as a function of $t_f$ are shown in Fig 3(b). The effective magnetization depends on saturation magnetization and surface anisotropy field ($H_s$) as,

$$\mu_0 M_{eff} = \mu_0 M_s - \mu_0 H_s \quad (3)$$

The dependence of $H_s$ and g-values on FeGaB thickness is shown in the inset of Fig. 3(b). Since the surface anisotropy field is defined as $H_s = 4K_s/(M_s t_f)$, where $K_S$ is the surface anisotropy constant, we have calculated the ratio $K_s/M_s = 0.168$ T.nm from the slope of the $H_s$ vs $1/t_f$ plot (Inset Fig. 3(b)). The surface anisotropy is maximum at the lower thickness presumably due to enhanced pinning of magnetization by surface defects and roughness of the surface. The anisotropy field drops to zero when $M_s$ becomes independent of the film thickness. We note that the g-factor which reflects the ratio of orbital and spin magnetic momenta has a reduced value compared to the g-factor of Fe or FeGa. However, we also see an enhancement in g-factor at lower film thicknesses, which is presumably due to degradation of magnetic properties near the interface [34]. It is well known that the g-factor is influenced by surface and interface effects as it depends on the local symmetry. Such effects may lead to strong enhancements of the ratio of orbital and spin momenta [35].

The enhanced role of surfaces on magnetization dynamics of these films is also seen in the behavior of the linewidth $\Delta H$. Fig 4(a) shows the dependence of $\Delta H$ on excitation frequency. These data have been analyzed in the framework of the Landau - Lifshitz - Gilbert (LLG) equation [36],

$$\Delta H = \Delta H_0 + \frac{4\pi f \alpha}{\gamma} \qquad (4)$$

where $\Delta H_0$ is the inhomogeneous line broadening caused by sample imperfections and surface defects, and $\alpha$ is the Gilbert damping coefficient. The linear frequency dependence of $\Delta H$ seen in Fig. 4(a) suggests that the magnetization damping is primarily of the LLG type with negligible contribution from two-magnon processes [37]. The $\alpha$ and $\Delta H_0$ are dominated by thin film size effects at lower thicknesses as seen in Fig. 4(b). The thickness dependence of $\alpha$ allows us to extract its surface ($\alpha_s$) and bulk ($\alpha_B$) components as [32, 38, 39]

$$\alpha = \alpha_B + \frac{\alpha_s}{t_f} \qquad (5)$$

A linear fit to the data as shown in the inset of Fig. 4(b) yields $\alpha_B = 2.18 \times 10^{-3}$ and $\alpha_s = 23.73 \times 10^{-3}$ nm$^{-1}$. The value of $\alpha_B$ is in excellent agreement with the reported data on FeGaB film [27].

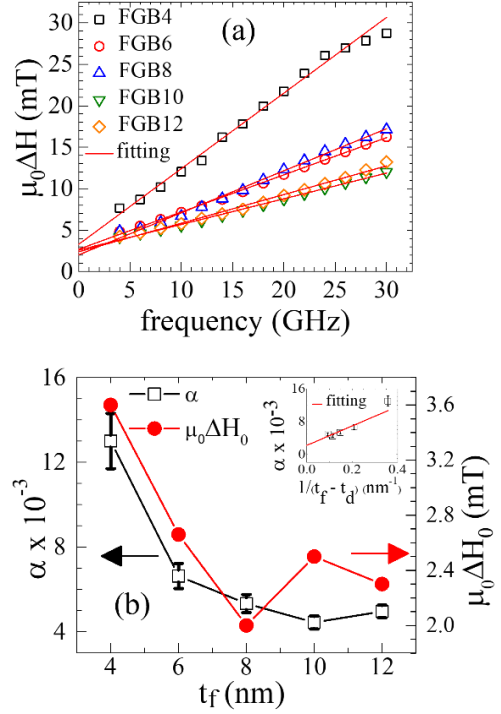

**Figure 4. (a)** Change in FMR linewidth ($\Delta H$) as a function of microwave frequency for FeGaB films of different thickness. Open symbols are calculated $\Delta H$ while red line is a LLG fitting. **(b)** Variation of the Gilbert damping constant ($\alpha$) and inhomogeneous linewidth broadening ($\Delta H_0$) as a function of film thickness. Inset shows a plot of $\alpha$ vs $1/t_f - t_d$ (where $t_d$ is the magnetic dead layer thickness) where open symbols are calculated values while red line is a linear fit to Eq. 5. Intercept at y-axis shows the $\alpha$ for bulk FeGaB system [27].

The large value of $\alpha_s$ together with the indication of a magnetic dead layer at the interface, suggest that the angular momentum of precessing magnetization is lost at the film – substrate interface. This inference is supported by our ISHE measurements, which are discussed next.

### C. Self-Induced ISHE and rectified voltage measurement

The flip – chip geometry used for the ISHE measurements is sketched in Fig. 1(b). The induced dc voltage is measured across the sample with its Y+ end connected to the positive terminal of the nanovoltmeter.

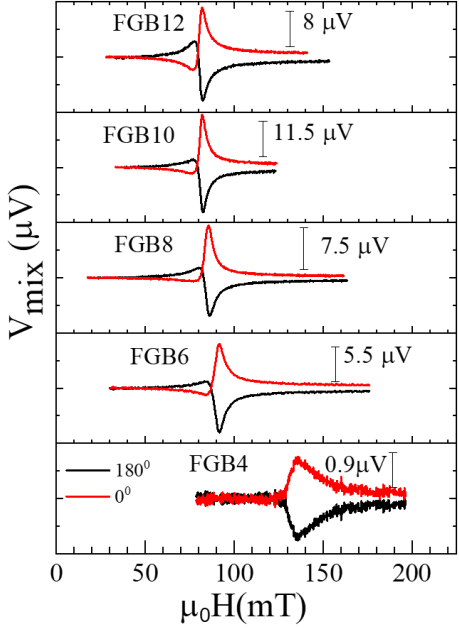

**Figure 5.** The measured dc voltage signal ($V_{mix}$) at 10 GHz for $\theta = 0^0$ and $\theta = 180^0$ angles. RF input power used for these measurements is +15dBm.

The in-plane external dc magnetic field is directed along the X-axis. Fig. 5 shows the variation of the output voltage of five films at 10 GHz excitation as the dc field is scanned across $H_r$. The noteworthy features of these data are: (i) the dc signal appears at the resonance, (ii) the signal is antisymmetric on field reversal, (iii) the line-shape is predominantly symmetric around $H_r$, (iv) the signal drops significantly as the film thickness is reduced and (v) the linewidth mimics the behavior of $\Delta H$ seen in FMR measurements.

The dc voltage generated across a thin film of metallic ferromagnet undergoing FMR derives contributions from several coherent and incoherent spin and charge scattering processes. For example, the RF component of the precessing magnetization results in a dc voltage due to AMR, and AHE in the magnetic layer through the **j** x **m** term in the force equation. Here **j** and **m** are RF current and magnetization respectively, which may not be in phase. The RF current in our measurement geometry comes from an inductive coupling between the waveguide and metallic film. The directions of rf current, rf magnetic field (h), dc field and voltage leads are sketched in Fig. 1(b). While the AMR related rectification effects in the geometry shown in the Fig. 1(b) may not contribute to any symmetric dc signal as the rf current is directed orthogonal to the dc voltage leads, it does give rise to an asymmetric voltage which varies as $cos2\Theta_H cos\Theta_H$ where $\Theta_H$ is the angle between dc magnetic field and rf current [20]. However, in the geometry of our experiment the AHE does contribute to a symmetric signal through the relation [40],

$$V_{AHE} = \frac{\rho_{AHE}.d.J_{rf}}{2}.m_{tz}.cos\Psi \qquad (6)$$

where $\rho_{AHE}$ is the anomalous Hall resistivity, d is the distance between the Hall contacts, $J_{rf}$ is the rf current density, $m_{tz}$ is the perpendicular component of the dynamic magnetization and $\Psi$ is the phase lag between rf current and dynamic magnetization. For both AMR and AHE related rectified voltages $\Psi$ is important in addition to the magnitudes of AMR and AHE in FeGaB films. We have measured the PHE, AMR and AHE in a 12 nm thick film of FeGaB patterned in a Hall bar geometry. Since these transport phenomena are characteristic features of spin – orbit interaction dominated electronic transport in magnetic alloys [41], their magnitude establishes the strength of spin – orbit coupling in FeGaB and their contribution to dc voltage generated under the FMR condition. The field symmetrized in-plane resistivity ($\rho_{xx}$) and PHE resistivity ($\rho_{xy}$) of a 12 nm thick film measured at 300 K are displayed in Fig.6 as a function of the angle $\phi$ between the magnetic field and transport current **I**.

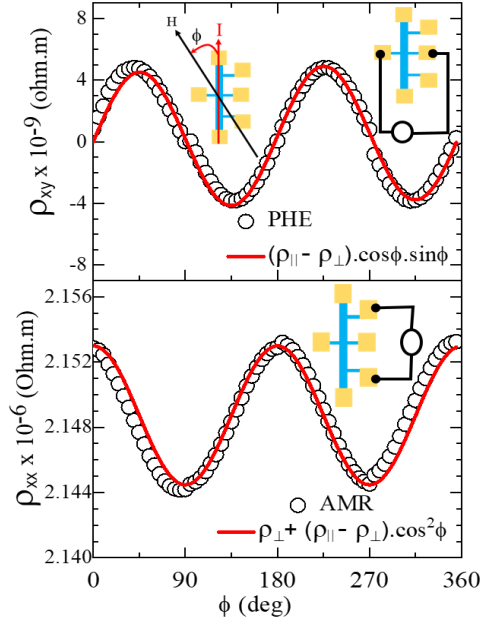

**Figure 6.** Angular dependence of resistivity $\rho_{xy}$ and $\rho_{xx}$ at $\mu_0 H = 1$ Tesla applied in the film plane. Red solid line is a fit to $(\rho_{II} - \rho_\perp) \cdot \cos\phi \cdot \sin\phi$ for PHE and $\rho_\perp + (\rho_{II} - \rho_\perp) \cdot \cos^2\phi$ for AMR. Inset shows the measurement geometry for AMR and PHE. The external dc field is in the plane of the film making angle $\phi$ with the direction of current (I). The field stays in the plane of the film when angle $\phi$ = changes from 0 to 360 degrees.

We note in Fig. 6 that the $\rho_{xx}$ and $\rho_{xy}$ follow the predicted angular dependence [42, 43] of the type $\cos^2\phi$ and $\cos\phi.\sin\phi$ respectively with, $\Delta\rho = (\rho_{II} - \rho_\perp) = 8.5 \times 10^{-9}$ Ohm.m. This yields a $\Delta\rho/\rho$ of $\approx 0.4$ %, which is five times smaller than the AMR seen in films of permalloy [41]. This reduced value of AMR suggests that the contribution of PHE to asymmetric signal will be much smaller compared to that in permalloy films.

The Hall resistivity ($\rho_{xy}$) of the film measured at ambient temperature ($\approx 300$ K) as a function of magnetic field is shown in Fig. 7. It follows the magnetic field dependence of the out of plane magnetization as shown in the inset of Fig. 7, and reaches a saturation value of (54 n$\Omega$ m) at $\approx 1.4$ T. The saturation field for $\rho_{xy}$ is the same as the saturation field for out-of-plane magnetization. The $\rho_{xy}$ for this FeGaB field is large compared to the AHE of permalloy ($\approx 0.96$ n$\Omega$ m). Such large Hall resistivity is also seen in several other metallic glasses [40, 44].

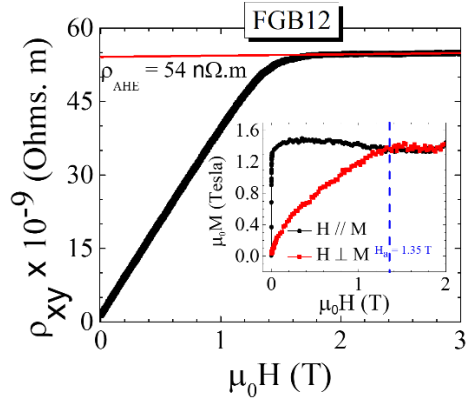

**Figure 7.** Hall resistivity ($\rho_{xy}$) measurement of FGB12 sample at room temperature. Inset shows the dc magnetization measurement of FGB12 sample in in-plane and out of plane geometry.

We have estimated the possible value of symmetric dc voltage generated by AHE in FGB12 sample using Eq. 6 assuming $\cos\psi = 1$ (Maximum), precession angle $\theta_C \approx 0.2^0$ and $J_{rf}$ is same ($\approx 1.15 \times 10^6$ A/m$^2$) as in the CPW signal line (See supplementary sec - C). This yields a peak value of 0.27 $\mu$V, which is $\approx 7$ % of the total symmetric signal shown in Fig. 8(e).

Here we must also mention that in a co-planar waveguide configuration the rf field would eventually become perpendicular to the film plane following the Ampere's Law. As shown in Ref. 20, the perpendicular component will yield $\Theta_H$ independent field asymmetric AHE voltage. This component of the rf field ($h_{rf}^\perp$) also produces a symmetric voltage due to PHE which does not change polarity on field reversal [45]. However, here we ignore the possible contributions of the out-of-plane rf field due to two reasons: one – the inductive rf current is expected to be localized in the region of the sample which is just above the signal line of the waveguide and therefore this region does not see much of the $h_{rf}^{\perp}$. Secondly, The PHE contribution of $h_{rf}^{\perp}$ should act against the dc signal produced by AHE on field reversal. However, unlike the behavior seen in Ref. 40, in our case the line shape and its magnitude remain the same when the polarity of the dc field is reversed.

The second source of dc voltage is the spin pumping effect of precessing magnetization. In our case, the magnetic dead layer at the film substrate interface may promote a unidirectional flow of spin current towards the substrate (See supplementary Fig. S2). This symmetry breaking process would result in a dc voltage through the inverse spin Hall effect in the FeGaB near the interface. The charge current produced by the ISHE is given as

$$\boldsymbol{J_c} = \theta_{SH}(\boldsymbol{J_s} \times \boldsymbol{\sigma}) \quad (7)$$

where $\theta_{SH}$ is the spin Hall angle which represents the ISHE efficiency in a material, $\boldsymbol{\sigma}$ is the spin polarization vector determined by the applied dc magnetic field and $\boldsymbol{J_s}$ is the spin current density. The polarity of the dc voltage measured in our experiment is consistent with Eq. 7 if it is assumed that the spin Hall angle of FeGaB is positive. Since the spin pumping process involves incoherent spins their leads to a voltage signal that is phase independent and symmetric around $H_r$.

We may also point out here that the rf current flowing through the dead layer may produce an alternating spin current because of SHE, which on entering the FM will affect the precession of magnetization and produce an antisymmetric rectified signal due to AMR. Finally, heating of the magnetic stripe by the absorbed RF power at resonance may produce temperature gradients across the film thickness, and hence a symmetric Nernst voltage.

We have decomposed the measured dc voltage into symmetric and anti-symmetric Lorentzian contributions [46, 47],

$$V_{mix} = K_s \frac{\delta H^2}{(H-H_r)^2+\delta H^2} + K_{as} \frac{-2\delta H(H-H_r)}{(H-H_r)^2+\delta H^2} \quad (8)$$

where $K_s$ and $K_{as}$ are symmetric and antisymmetric coefficients and $\delta H$ is the half width at half maxima.

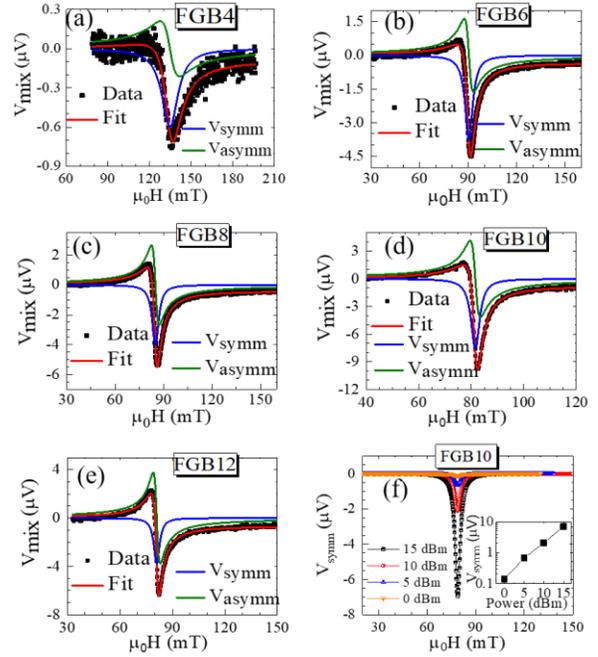

**Figure 8. (a-e)** Shows the fitting of experimental data for the samples FGB4, FGB6, FGB8, FGB10 and FGB12 using Lorentzian symmetric and antiymmetric equation at $\theta = 180^0$ where $V_{mix} = V_{symm} + V_{asymm}$. **(f)** Power dependent symmetric part ($V_{symm}$) for the sample FGB10 at 10 GHz, inset shows the increase of maximum voltage as a function of input RF power at 10 GHz.

The result of this analysis for five films of different thicknesses is shown in Fig. 8 (a through e). In Fig. 8(f) we show the symmetric component of the voltage of 10 nm thick FeGaB film extracted from the measurements $V_{dc}$ performed at different RF power.

A linear power dependence confirms that the data reported here have been collected in the linear regime of response. In order to ensure that the measured dc signal is not an artifact of sample heating, we have measured the $V_{mix}$ signal of the 10 nm thick film at different scan rates of the dc field. The slower scan rate amounts to a larger dwell time near $H_r$. However, even an order of magnitude change in the scan rate does not affect the position or amplitude the $V_{mix}$ voltage. An unperturbed value of resonance field when measurements were performed at different rf power further confirm absence of sample heating in these experiments. (See supplementary Fig. S4). The contributions of various spin rectification effects to the symmetric and antisymmetric part of the measured dc voltage are discussed in supplementary sec- E. We have estimated these contributions for FGB12 sample. A similar analysis can be carried out for thinner films as well with the knowledge of their AHR.

We now quantify the weight of symmetric ($W_s$) and antisymmetric ($W_{as}$) parts of the dc voltage generated in the FeGaB thin films in terms of coefficient $K_s$ and $K_{as}$ as $W_s = (K_s/ (K_s + K_{as}))$ and $W_{as} = (K_{as}/ (K_s + K_{as}))$ [46]. The variation of $W_s$ and $W_{as}$ as functions of film thickness is shown in Fig. 9. It is interesting to note that the symmetric weight decreases while antisymmetric weight increases with increase in thickness of FeGaB. This behavior can be understood as follows: (i) The symmetric contribution comes mainly due to the SOC in FeGaB, which is larger in thinner films as suggested by the variation of the g-factor shown in the inset of Fig. 3. Also, the rectification contribution of AHE may be larger as the $\rho_{xy}$ scales with $\rho_{xx}$ which is higher for thinner films due to size effects. (ii) The antisymmetric contribution is due to spin rectification effects derived from PHE, is expected to be larger in thicker films as they allow stronger inductive currents due to their high electrical conductivity (see Fig. 9 inset).

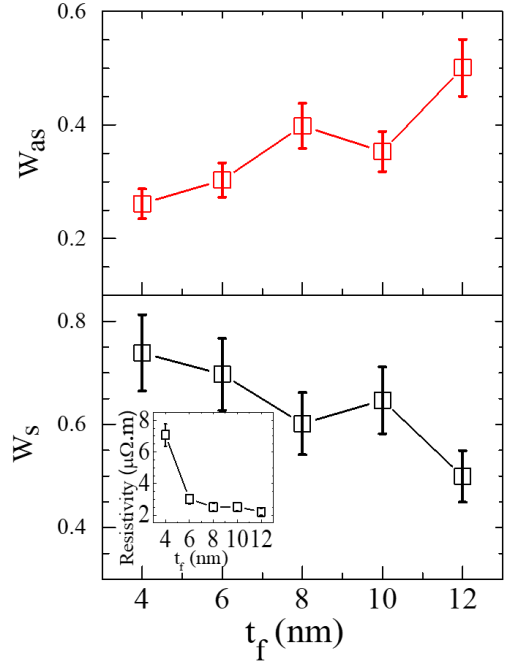

**Figure 9.** Symmetric ($W_s$) and antisymmetric weight ($W_{as}$) of the DC signal at f = 10 GHz plotted as a function of FeGaB thickness. $W_s = K_s/(K_s + K_{as})$ and $W_{as} = K_{as}/(K_s + K_{as})$. Inset shows the variation of resistivity of FeGaB as a function of film thickness.

## IV. Summary

We have measured the induced dc voltage in amorphous thin films of $Fe_{78}Ga_{13}B_9$ alloy when their magnetization is driven to resonant precession on excitation with a microwave field of variable frequency. This voltage, measured orthogonal to the direction of the dc magnetic field, is antisymmetric on field reversal and its line shape has both symmetric and asymmetric components centered around the FMR resonance field. The measurements of static magnetization and frequency dependent FMR on films of different thicknesses indicate a dominant role of thin film size effects in setting the values of $M_s$, FMR linewidth, Gilbert damping parameter and g-factor. The thin film size effects become increasingly stronger as the film thickness drops below ≈ 8 nm. The thickness dependence of these static and dynamic magnetization parameters also indicates the formation of a ≈ 1.2 nm thick magnetic dead layer at the film – substrate interface. We have considered several processes that may lead to an ISHE - like voltage at FMR in these films. The RF power and dwell-time dependence of $V_{dc}$ and dP/dH rule out any contribution of thermally driven spin currents in producing the $V_{dc}$. The symmetric component of $V_{dc}$ and its polarity, on the other hand, suggest a preferential flow of spin currents produced by precessing magnetization towards the film – substrate interface. We have also estimated the contribution of anomalous Hall effect to this dc voltage. The $V_{dc}$ appears to be a compound effect of ISHE in the FeGaB film and its symmetry breaking dead-layer together with spin rectification effects of AHE. Our experiments suggest that the FMR induced $V_{dc}$ in plain films of high AHE amorphous alloys needs to be considered for correct interpretation of the ISHE data on spin orbit coupled normal metal – ferromagnet heterostructures.

**Acknowledgements**

This research is supported by the Air Force Office of Scientific Research, Grant # FA9550-19-1-0082.